# 2D van der Waals Nanoplatelets with Robust Ferromagnetism

Michael C. De Siena,[1,‡] Sidney E. Creutz,[1,‡,||] Annie Regan,[1,§] Paul Malinowski,[2] Qianni Jiang,[2] Kyle T. Kluherz,[1,4] Guomin Zhu,[3,4] Zhong Lin,[2] James J. De Yoreo,[1,3,4] Xiaodong Xu,[2,3] Jiun-Haw Chu,[2] and Daniel R. Gamelin[1]*

[1]*Department of Chemistry, University of Washington, Seattle, WA 98195, USA*
[2]*Department of Physics, University of Washington, Seattle, WA 98195, USA*
[3]*Department of Materials Science and Engineering, University of Washington, Seattle, WA 98195, USA*
[4]*Physical Sciences Division, Pacific Northwest National Laboratory, Richland, WA 99352, USA*

*Corresponding author's e-mail: gamelin@uw.edu

**Abstract.** We have synthesized unique colloidal nanoplatelets of the ferromagnetic two-dimensional (2D) van der Waals material $CrI_3$ and have characterized these nanoplatelets structurally, magnetically, and by magnetic circular dichroism spectroscopy. The isolated $CrI_3$ nanoplatelets have lateral dimensions of ~25 nm and ensemble thicknesses of only ~4 nm, corresponding to just a few $CrI_3$ monolayers. Magnetic and magneto-optical measurements demonstrate robust 2D ferromagnetic ordering in these nanoplatelets with Curie temperatures similar to those observed in bulk $CrI_3$, despite the strong spatial confinement. These data also show magnetization steps akin to those observed in micron-sized few-layer 2D sheets and associated with concerted spin-reversal of individual $CrI_3$ layers within few-layer van der Waals stacks. Similar data have also been obtained for $CrBr_3$ and anion-alloyed $Cr(I_{1-x}Br_x)_3$ nanoplatelets. These results represent the first example of laterally confined 2D van der Waals ferromagnets of any composition. The demonstration of robust ferromagnetism at nanometer lateral dimensions opens new doors for miniaturization in spintronics devices based on van der Waals ferromagnets.

Long-range magnetic order has recently been demonstrated in micron-sized two-dimensional (2D) magnetic semiconductors,[1-2] opening opportunities for atomically thin spintronics.[3-5] For high-density devices, however, a fundamental limitation is the instability of ferromagnetism as lateral sizes approach the nanoscale.[6-7] It is thus of both fundamental and technological importance to explore the magnetic properties of 2D layered magnets with strong lateral confinement. Here, we report successful synthesis of unique colloidal nanoplatelets of 2D magnets $CrX_3$ (X = I, Br) and their alloys. The miniaturization and associated solution compatibility demonstrated here suggest new opportunities for studying 2D spin effects at the nanoscale and exploiting nanostructuring in high-density spintronics.

An interesting aspect of the ferromagnetism in $CrI_3$ is that the ground-state ($^4A_2$) single-ion anisotropy (*D*) of its pseudo-octahedral $Cr^{3+}$ constituents is negligible,[8] yet the Curie temperature is still relatively high ($T_C$ ~ 45 K in monolayers). The magnetic anisotropy underlying $T_C$ comes entirely from the 2D morphology of the individual layers, and specifically from the presence of in-plane (anisotropic) ferromagnetic superexchange interactions that pin the spin orientation of any given $Cr^{3+}$ and generate a gap in the spin-wave density of states.[8-9] As the lateral dimensions of the $CrI_3$ plane are decreased, this morphological stabilization should diminish due to broken symmetry at the edges, eventually causing the ferromagnetism to be surmountable by thermal fluctuations and the $CrI_3$ to become superparamagnetic. For example, ferromagnetic *hcp* cobalt metal has a very high $T_C$ of >1000 K in bulk but becomes superparamagnetic at room temperature in nanocrystals.[10]

Theory predicts that the electronic structures of $CrI_3$ nanoribbons should be strongly dependent on their edge structures, displaying quantum size effects in the energy gap and new edge states around the Fermi level.[11] To date, however, there have been no experimental examples of laterally confined 2D magnets of any type. It remains an open question whether 2D ferromagnetism can survive lateral truncation of the 2D structural periodicity.

Here we report the bottom-up synthesis and physical characterization of colloidal few-layer $CrX_3$ (X = I, Br) nanoplatelets with average lateral dimensions of only ~25 nm, focusing on $CrI_3$. Common solution routes to colloidal metal-halide nanocrystals[12] frequently involve high-temperature reaction of halide sources (*e.g.*, alkylammonium halides, benzoyl halides, trimethylsilyl halides) with simple metal salts (*e.g.*, metal acetates, carbonates) solubilized by surfactants (*e.g.*, long-chain carboxylates, amines). In our hands, attempts to synthesize $CrX_3$ *via* similar approaches using simple Cr(III) precursors and common surfactants did not work (see Supporting Information). Octahedral Cr(III) is canonically substitutionally inert,[13] making these simple Cr(III) precursors difficult to solubilize and relatively unreactive. We therefore sought more reactive precursors that would be highly soluble in nonpolar, noncoordinating solvents such as alkanes or toluene, and that would also react in a way that produces only innocuous and readily separable byproducts. We found that $Cr(OR)_3$ (R = 1,1-di-*t*-butylethoxide) meets all of these requirements. Similarly, trimethylsilyl iodide (TMSI) was identified as a reactive and alkane-soluble iodide precursor.[14] When an anhydrous toluene solution of $Cr(OR)_3$ with excess TMSI is immersed with vigorous stirring into an oil bath preheated to 135 °C, abrupt formation of a black precipitate is observed after ~5 min. The rapid onset of nucleation under these conditions is well-suited to the formation of high-quality nanocrystals.[15] Because no solubilizing ligands are present, the resulting nanocrystals precipitate from solution rather than forming a colloidal dispersion. They can be separated from the clear supernatant by centrifugation and washing with hexanes. The nanocrystals can then be resuspended in dichloromethane through vigorous ultrasonication to generate a clear, dark-green solution (Fig. 1a).

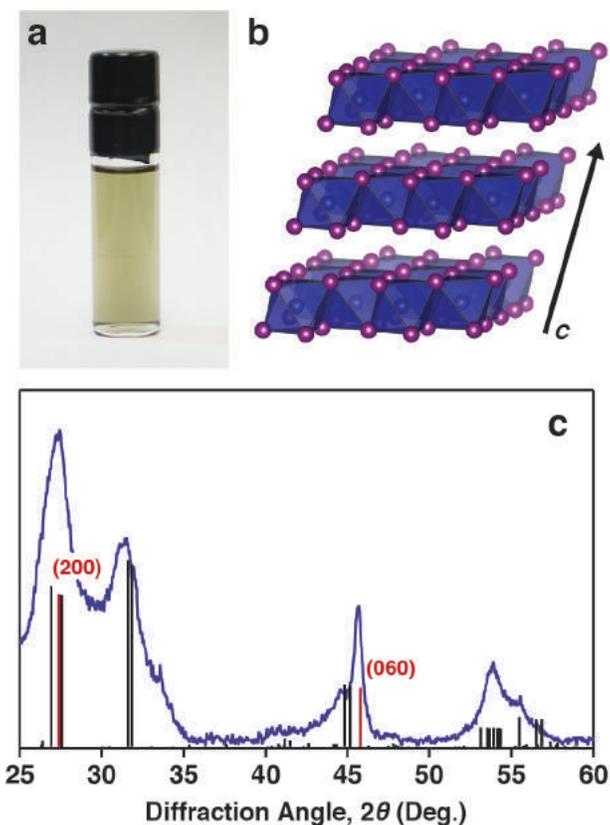

**Figure 1. (A)** Photograph of a solution of $CrI_3$ nanocrystals in DCM. **(B)** Illustration of the room-temperature crystal structure of bulk $CrI_3$, showing a section of three monolayers with the nearly ABC stacking arrangement of the monoclinic $C2/m$ structure.[18] The arrow indicates the direction of the crystallographic $c$-axis. **(C)** Powder X-ray diffraction data for $CrI_3$ nanocrystals (blue) measured with Cu $K_α$ radiation, compared to a reference pattern for $CrI_3$ ($C2/m$, ICSD Coll. Code 251654).[18] Prominent reflections from planes parallel to the $c$-axis are highlighted in red and labeled with the Miller indices of the corresponding planes.

Interlayer magnetic coupling in $CrI_3$ is tightly connected to the layer stacking arrangement.[16-17] Figure 1b illustrates the lattice structure of bulk $CrI_3$, which adopts a monoclinic $C2/m$ structure at room temperature and transitions to a rhombohedral $R$-3 phase below 220 K.[18] Figure 1c shows powder X-ray diffraction (PXRD) data collected for the isolated nanocrystals. Although high sensitivity to air complicated the PXRD measurements (see Methods), these data confirm the identity, structure, and morphology of the isolated material as $CrI_3$, best matching the $C2/m$ monoclinic phase. In this phase, stacked $CrI_3$ monolayers are displaced along the $a$-axis with the nearly ABC arrangement[18] associated with antiferromagnetic interlayer coupling.[19-20] Notably, substantial Scherrer broadening is observed in Fig. 1c for reflections from planes parallel to the *ab*-plane. Only reflections from planes lying mostly or entirely parallel to the *c*-axis are reasonably narrow. This result suggests substantial crystalline shape anisotropy. As described below, these nanocrystals are in fact $CrI_3$ nanoplatelets with lateral dimensions ~5 times greater than their few-layer thicknesses. The (060) reflection at $2\theta = 46°$ is particularly narrow, and its width can be used to estimate a mean lateral nanoplatelet size of ~20 nm for the sample of Fig. 1. Notably, these PXRD peak widths and the colloidal stability of these nanocrystals preclude the existence of large bulk-like crystals within this ensemble.

Transmission electron microscopy (TEM) was used to further characterize the $CrI_3$ nanocrystals and the results are summarized in Fig. 2. Figure 2a shows an overview TEM image of an ensemble of $CrI_3$ nanocrystals. The absence of surface ligands causes the nanocrystals to aggregate when cast onto the TEM grids, making it difficult to identify many well-isolated nanocrystals. Figure 2b shows a selected-area electron diffraction (SAED) image of the same aggregated nanocrystals. Its integrated profile is shown in Fig. 2c compared to the calculated reference lines for $CrI_3$ (*C2/m*). These data confirm the identity of these nanocrystals as monoclinic $CrI_3$.

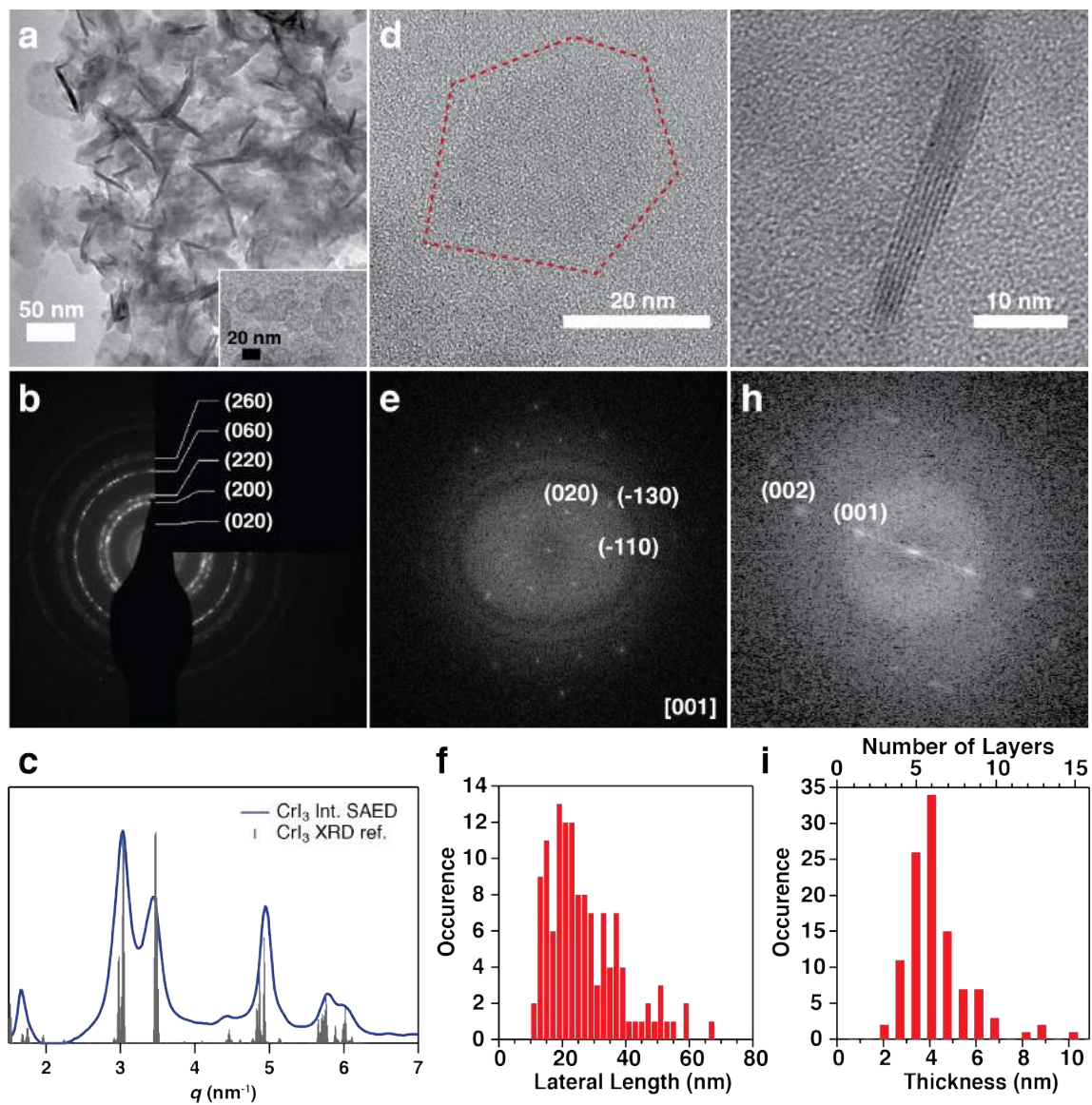

**Figure 2.** **(A)** TEM image of an aggregate of $CrI_3$ nanoplatelets. Inset: Zoomed in view of a string of isolated nanoplatelets. **(B)** Selected area electron diffraction (SAED) image of aggregated $CrI_3$ nanoplatelets. The rings are indexed to their corresponding diffraction planes. **(C)** Integrated profile of the SAED diffraction rings, compared to the calculated reference lines for $CrI_3$ ($C2/m$). **(D)** High-resolution TEM image of a single isolated nanoplatelet. **(E)** Fourier transform of the image from panel D, showing peaks consistent with the nanoplatelet being a single crystalline domain with the monoclinic $C2/m$ structure along the [001] zone axis. **(F)** Distribution of $CrI_3$ nanoplatelet lateral sizes from a survey of over 100 nanoplatelets, yielding an ensemble width of 26 ± 11 nm. **(G)** Edge-on HRTEM image of a single stack of nanoplatelets. **(H)** Fourier transform of the image from panel G, showing peaks consistent with van der Waals stacked $CrI_3$ nanoplatelets with the $C2/m$ structure. **(I)** Distribution of $CrI_3$ nanoplatelet thicknesses from a survey of over 100 nanoplatelets. The average nanoplatelet thickness is 3.7 ± 0.7 nm, corresponding to 6 ± 1 $CrI_3$ monolayers. All data were collected at room temperature.

The inset to Fig. 2a shows a low-resolution image of several individual nanocrystals side-by-side. These nanocrystals appear disk-like and have low contrast, consistent with thin nanoplatelets. Figure 2d shows a high-resolution image of a single nanoplatelet. Although faint, faceting consistent with hexagonal $CrI_3$ is observable. Careful inspection further reveals lattice fringes in this image. Figure 2e plots the Fourier transform of this image, revealing distinct peaks characteristic of single-crystal $CrI_3$ (*C*2/*m*). The sizes of over 100 individual nanoplatelets were measured, yielding an ensemble width of 26 ± 11 nm (Fig. 2f). The smallest nanoplatelets identified were only 10 nm across, and the largest nanoplatelet reached 66 nm across.

Several nanocrystals can be seen lying on their sides, resulting in thin dark stripes in Fig. 2a. Figure 2g shows a high-resolution TEM image of one isolated edge-on structure, in which six parallel thinner lines are observed. Figure 2h plots the Fourier transform of this image, revealing a series of spots that can be indexed to (00*n*) peaks. These data show an interlayer spacing of 0.69 ± 0.07 nm, consistent with the value of 0.66 nm expected from the $CrI_3$ (*C*2/*m*) crystal structure. Over 100 edge-on nanoplatelets were surveyed, from which an average thickness of 3.7 ± 0.7 nm is obtained (Fig. 2i), corresponding to only 6 ± 1 $CrI_3$ monolayers. Similar results were obtained for $CrBr_3$ nanoplatelets prepared analogously (see Supporting Information). Beyond $CrI_3$ and $CrBr_3$, this synthetic approach also allows formation of anion-alloyed $Cr(I_{1-x}Br_x)_3$ nanoplatelets, providing a mechanism for continuous tuning of the nanoplatelet optical and magnetic properties (see Supporting Information).

The electronic structures and magnetism of $CrBr_3$ and $CrCl_3$ single crystals have previously been investigated by transmission and reflection magnetic circular dichroism (MCD) spectroscopies,[21-22] including down to a single monolayer.[2,23-24] Figure 3 summarizes the 5 K absorption (extinction) and transmission MCD spectra of representative $CrBr_3$ and $CrI_3$ nanoplatelet ensembles. MCD spectra are plotted as the differential absorbance of left and right circularly polarized light (ΔA) normalized to the absorbance at the first peak maximum ($A_{max}(^4T_2)$) after accounting for scattering.

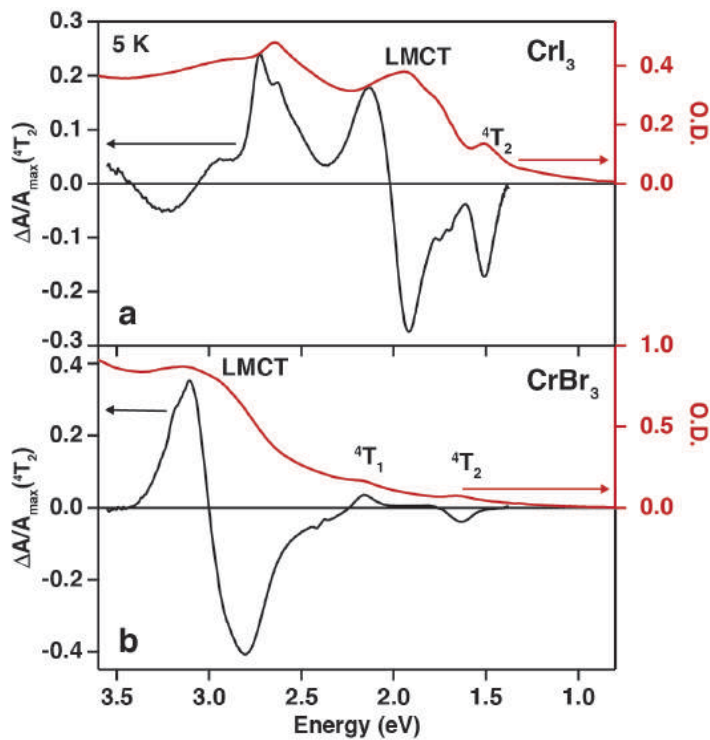

**Figure 3.** Transmission magnetic circular dichroism (MCD) spectra of **(A)** $CrI_3$ and **(B)** $CrBr_3$ nanoplatelets measured at 5 K and at a magnetic field of 5 T, compared with the 5 K zero-field electronic absorption (extinction) spectra. Absorption axes are labeled by optical density (O.D.). MCD spectra are plotted as $\Delta A/A_{max}(^4T_2)$, *i.e.*, the differential absorbance of left and right circularly polarized light (ΔA) normalized to the absorbance at the first peak maximum ($A_{max}(^4T_2)$) after accounting for the scattering baseline. The assignments of select optical transitions are indicated.

The $CrI_3$ nanoplatelet spectra (Fig 3a) are better understood by first examining the spectra of $CrBr_3$ nanoplatelets (Fig. 3b). As in bulk $CrBr_3$,[21-22] the $CrBr_3$ nanoplatelet spectra show two weak absorption features at low energies (1.6, 2.2 eV) attributable to the $^4A_2 \rightarrow {}^4T_2$ and $^4T_1$ ligand-field transitions of pseudo-octahedral $[CrBr_6]^{3-}$. Each band is associated with an MCD feature in which one polarization dominates. To higher energy, a pair of more intense π-type ligand-to-metal charge-transfer (LMCT) transitions appears, centered at ~2.75 eV and with opposite MCD polarizations.[25-26]

The absorption spectrum of the $CrI_3$ nanoplatelets (Fig. 3a) is very similar to the differential reflection spectrum of bulk $CrI_3$, but it differs slightly from that reported for monolayer $CrI_3$,[27] possibly due to interface effects in the latter. In Fig. 3a, the $^4T_2$ band remains the lowest-energy transition, but the LMCT transitions have shifted down to ~2.1 eV. A redshift of ~1.1 eV is predicted from the electronegativity difference between $Br^-$ and $I^-$,[28] in good agreement with the data. To higher energies in the $CrI_3$ spectra, several additional absorption and MCD features are observed, but overlap between LMCT and ligand-field transitions complicates band assignments. Both $CrI_3$ and $CrBr_3$ thus have their absorption gaps determined by the same highly localized $^4A_2 \rightarrow\ ^4T_2$ $Cr^{3+}$ ligand-field transition. Notably, the MCD rotational strength of the $^4T_2$ excitation is substantially (~5x) greater in the $CrI_3$ nanoplatelets than in the $CrBr_3$ nanoplatelets, likely reflecting enhanced configuration interaction with the strongly optically active LMCT states in $CrI_3$ due to their lower energies.

Variable-temperature and variable-field MCD measurements were performed to assess the magnetism of these $CrI_3$ and $CrBr_3$ nanoplatelets. Figure 4a plots $CrI_3$ MCD spectra collected at several temperatures between 5 and 200 K (see Supporting Information for $CrBr_3$ data). The inset to Fig. 4a plots the MCD amplitudes extracted from these data, from which $T_C$ = 54 K is determined. This value is smaller than in bulk ($T_C$ = 61 K),[18] but the trend is consistent with the decrease to $T_C$ = 45 K reported for monolayer $CrI_3$ sheets.[2] Figure 4b plots variable-field MCD spectra of the $CrI_3$ nanoplatelets collected at 5 K. All of the spectral features show the same field dependence, confirming their common origin. The inset to Fig. 4b plots the $CrI_3$ nanoplatelet

MCD intensity as a function of applied field, revealing a sizable hysteresis that confirms retention of ferromagnetism in these nanoplatelets. Multiple distinct inflections are observed at intermediate fields during the field sweep (see Supporting Information).

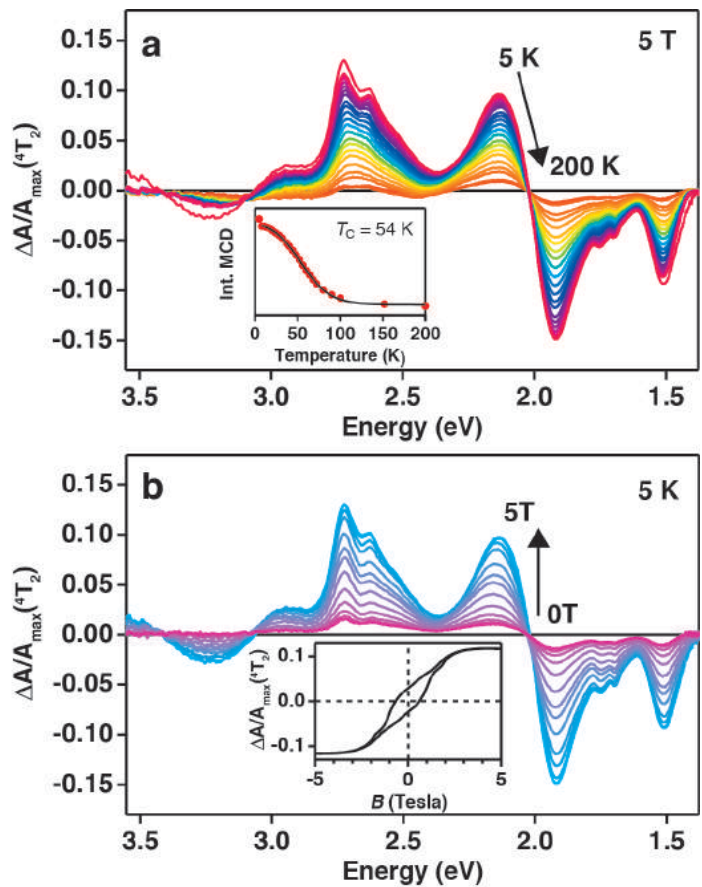

**Figure 4. (A)** Variable-temperature (5 to 200 K) MCD spectra of $CrI_3$ nanoplatelets measured at 5 T. Inset: Plot of integrated absolute 5 T MCD intensity as a function of temperature, yielding a Curie temperature of ~54 K. The solid curve in the inset is a guide to the eye. **(B)** Variable-field (0 - 5T) MCD spectra of $CrI_3$ nanoplatelets measured at 5 K. Inset: Magnetization *vs* magnetic field plot for $CrI_3$ nanoplatelets at 5 K, as probed using the MCD signal at 455 nm. The sample was first magnetized at 5 T then swept to -5 T and back to 5 T during data collection.

Figures 5a,b compare the MCD hysteresis curve from Fig. 4c with magnetic hysteresis data for the same nanoplatelets collected by vibrating sample magnetometry (VSM) on a slightly expanded *x*-axis scale. Although less pronounced, several inflections are also observed in the VSM data. Interestingly, the fields at which these inflections occur coincide almost *exactly* with the fields at which magnetization steps are observed in mechanically exfoliated multilayer $CrI_3$ sheets. To illustrate, the shaded grey bars in Fig. 5 mark the fields at which abrupt magnetization steps occur in four-layer $CrI_3$ sheets with micron lateral dimensions when magnetized along their easy axis.[23] The steps in the nanoplatelet data appear less abrupt due to orientation averaging because of more gradual magnetization when the external field (*B*) is not along the nanoplatelet easy axis. These steps have been assigned to magnetization reversal of individual $CrI_3$ monolayers within multilayer stacks.[23] We thus attribute the inflections observed in Fig. 5a,b to magnetization reversal of individual monolayers within multilayer $CrI_3$ nanoplatelets, as depicted schematically in Fig. 5d.

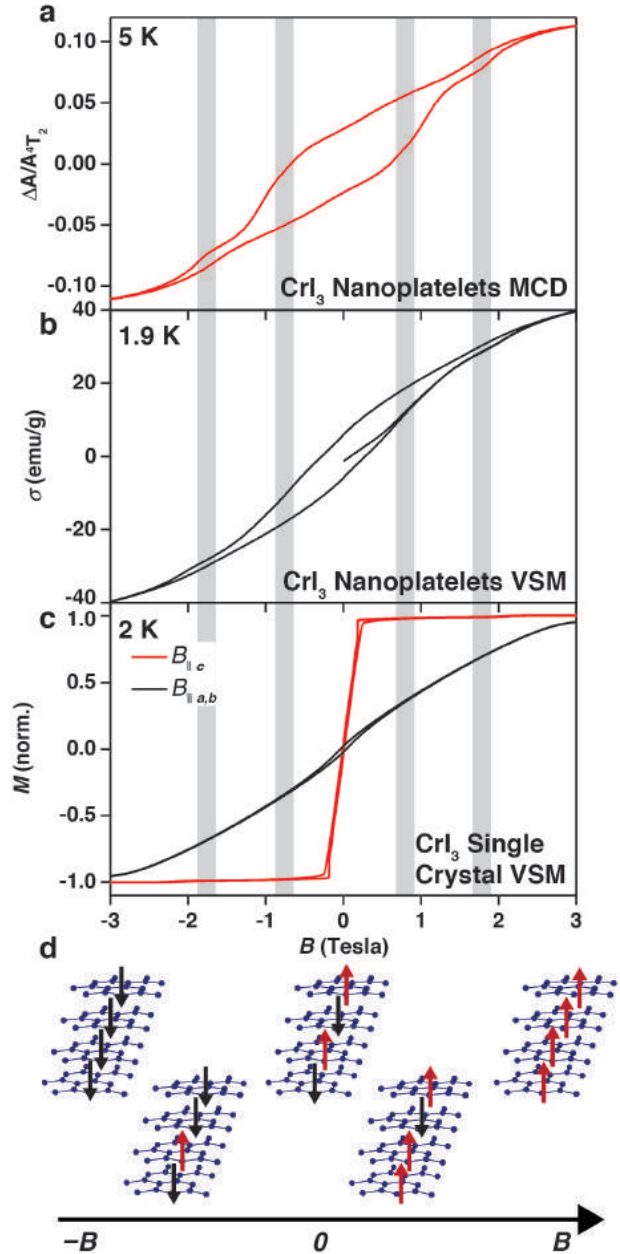

**Figure 5.** Magnetization *vs* field plots measured by **(A)** MCD spectroscopy and **(B)** vibrating sample magnetometry (VSM) on the same sample of randomly oriented $CrI_3$ nanoplatelets. **(C)** VSM data from a $CrI_3$ single crystal aligned both parallel and perpendicular to the applied field are also plotted on the same field axis. The vertical gray bars represent the fields at which the spins of individual layers of a four-layer $CrI_3$ sheet have been reported to flip.[23] **(D)** Schematic of magnetic structures of a representative four-layer $CrI_3$ nanoplatelet. Purple spheres represent $Cr^{3+}$ ions and $I^-$ ions are omitted. The fields at which transitions between these magnetic arrangements occur are determined by the weak antiferromagnetic interlayer coupling.

The monolayer spin-flip events observed here are signatures of few-layer $CrI_3$. In these nanoplatelets, however, the fraction of edge $Cr^{3+}$ ions is vastly greater than in exfoliated $CrI_3$ sheets: Idealizing the nanoplatelets as hexagons, the average $CrI_3$ nanoplatelet here contains ~2360 $Cr^{3+}$ ions, of which ~10% occupy edge sites. Because of broken translational symmetry, these edge spins are not pinned with the same energies as the core spins, leading to spin canting or lower-temperature spin fluctuations. Lateral confinement should thus reduce the barrier to magnetization reversal, but the data here do not show such a reduction. From the magnetocrystalline anisotropy of bulk $CrI_3$,[29] we estimate that the barrier to monolayer magnetization reversal will not drop below the bulk value of $k_B T_C$ until lateral dimensions reach < ~6.5 nm (~230 $Cr^{3+}$ ions, ~34% edge sites, see Supporting Information), consistent with the bulk-like $T_C$ observed in Fig. 4a.

The nanoplatelets reported here represent a fundamentally new morphology for this emergent class of materials. Few-layer van der Waals ferromagnets like $CrI_3$ have previously only been prepared using mechanical exfoliation of bulk crystals. Remarkably, despite their extremely small (~25 nm) lateral dimensions, these $CrI_3$ nanoplatelets still show the same robust intra-layer ferromagnetism, interlayer antiferromagnetic coupling, and discrete layer-by-layer spin-flip magnetization as observed in mechanically exfoliated $CrI_3$ sheets with much larger dimensions. Moreover, $T_C$ appears undiminished by their very small volumes. This miniaturization, in conjunction with the unique solution processability of this form of $CrI_3$, opens interesting avenues for constructing fundamentally new types of programmable 2D magnetic quantum materials, for example, through their integration as localized "spin bubbles" within otherwise non-magnetic 2D heterostructures, through self-assembly into spin-bubble superlattices, through nucleation of lateral heterostructures, or even through combination with other solution-phase materials to construct free-standing nanoscale 2D van der Waals heterostructures with magnetic functionality. More generally, the demonstration that robust ferromagnetism persists in nanoscale $CrI_3$ indicates the possibility of high-density spintronic devices based on nanopatterned 2D ferromagnetic semiconductors.

**Methods**

**General considerations.** Unless otherwise stated, all measurements and synthetic manipulations were performed using standard Schlenk techniques under a dinitrogen atmosphere, or in a glovebox under an atmosphere of purified dinitrogen. Anhydrous tetrahydrofuran (THF), dichloromethane (DCM), ethyl ether, pentane, and toluene were purified through an alumina column pressurized with argon. Hexanes was further dried over sodium benzophenone and distilled before use. Xylenes was dried by refluxing over calcium hydride and distilled before use. All solvents were stored over 4Å sieves.

**Chemicals.** Unless otherwise stated, all chemicals were used as received without further purification. Zinc metal, chromium metal (99.995%), anhydrous $CrCl_3$, and $I_2$ (99.99%) were purchased from Alfa Aesar. Hexamethylacetone, methyllithium (1.6 M in ether), trimethylsilyl bromide (97%), and trimethylsilyl iodide (97%) were purchased from Sigma Aldrich. $Cr(OCMe^tBu_2)_3$ was synthesized as previously reported.[30]

**Synthesis of $CrX_3$ nanoplatelets.** The chromium precursor, $Cr(OCMe^tBu_2)_3$ (20 mg, 0.04 mmol) is dissolved in 2 mL of toluene (X = I) or xylenes (X = Br) in a 25 mL Schlenk tube equipped with a stir bar, giving a light blue solution. A total of 0.70 mmol of trimethylsilyl halide (bromide, iodide, or a mixture of both) is added to the reaction mixture. No observable change occurs. The Schlenk tube is sealed and immersed into a pre-heated oil bath at 135˚C (180°C for CrBr$_3$) with rapid stirring. After approximately 5 min, the solutions abruptly turned black and opaque, indicating the formation of the CrI$_3$ nanoplatelets as a precipitate. Heating was continued for an additional 10 min, and then the reaction was removed from the heating bath and allowed to cool to room temperature. The precipitate was separated from the colorless supernatant by centrifugation and washed three times with hexanes (5 mL). The precipitate was finally collected as a suspension in hexanes.

**Synthesis of CrI$_3$ single crystals.** Single crystals of CrI$_3$ were grown by chemical vapor transport using iodine as a self-transport agent following a procedure adapted from the literature with modification.[18] Cr(0) pieces and solid crystalline I$_2$ were loaded into a quartz tube and sealed under an evacuated argon atmosphere. The quartz tube was 10 cm long, inner diameter 14 mm, and outer diameter 16 mm, and the amount of loaded iodine was determined by ensuring that the pressure inside of the tube reached a value near atmospheric pressure upon reaching the highest growth temperature. The transport of material was achieved using the natural temperature gradient of an open-ended horizontal furnace. The source end of the tube is placed in the hot end of a 650/550°C temperature gradient and allowed to dwell for 7 days, and then allowed to slowly cool to room temperature. Crystals grew at the cold end of the tube as large, shiny black plates.

**Powder X-ray diffraction (XRD) measurements.** Samples were prepared for powder XRD by drop-casting suspensions of nanoplatelets from hexanes onto silicon wafers, and protecting them from air by sealing under Kapton film. Data were collected using a Bruker D8 Discover diffractometer.

**Magnetic circular dichroism (MCD) and absorption (extinction) measurements.** Samples for transmission MCD measurements were prepared by mixing dried nanoplatelets in polydimethylsiloxane (PDMS, viscosity 1,000 cSt). This mixture was then sandwiched between two quartz discs to make a mull suspension. Low-temperature magnetic circular dichroism (MCD) spectra were conducted with the samples placed in a superconducting magneto-optical cryostat (Cryo-Industries SMC-1659 OVT) oriented in the Faraday configuration. Samples were loaded into the cryostat under helium gas to minimize air exposure and sample decomposition. At liquid helium temperatures, the sample was screened for depolarization by matching the CD spectra of a chiral molecule placed along the optical path before and after the sample. Depolarization by the samples was less than 5% in each case. Transmission MCD spectra were collected using an Aviv 40DS spectropolarimeter. UV-Vis absorption (extinction) measurements were performed on similar mulls using an Agilent Cary 5000 spectrophotometer, and sample cooling was achieved using a flow cryostat with a variable-temperature sample compartment.

**Vibrating sample magnetometry (VSM) measurements.** A Quantum Design PPMS DynaCool was used for VSM measurements. Powders were loaded into plastic VSM powder sample holders and the single crystal was affixed to the end of a quartz paddle with varnish (VGE 7031). The paddle was then snapped into the VSM brass sample holder with another quartz paddle placed symmetrically above the sample to minimize the background coming from the quartz.

**Transmission electron microscopy (TEM) measurements.** TEM samples were prepared by drop casting suspensions of nanocrystals onto 400 mesh carbon-coated copper grids purchased from TED Pella, Inc. and dried under an inert atmosphere. Nanocrystal suspensions were prepared by ultrasonication of the materials in dichloromethane at a sonicator frequency of 20 kHz under an inert atmosphere. In a glovebox, TEM grids were loaded into a vacuum transfer holder to prevent sample exposure to air. TEM images were obtained on an FEI Titan microscope operated at 300 kV or on an FEI Tecnai F20 microscope operated at 200 kV. FFT images were generated using ImageJ, and brightness was adjusted to aid visualization.[31]

**Energy-dispersive X-ray spectroscopy (EDS) measurements.** For EDS analysis of nanoplatelet compositions, samples were drop cast onto silicon substrates and coated with a ~200 nm thick layer of carbon; spectra were acquired in an FEI Sirion Scanning Electron Microscope operating at 30 kV using an Oxford EDS spectrometer. Standardless quantification was used.

**Acknowledgments**
The development, structural and analytical characterization, and magnetic/magneto-optical characterization of these 2D ferromagnets were all supported as part of Programmable Quantum Materials (Pro-QM), an Energy Frontier Research Center funded by the U.S. Department of Energy (DOE), Office of Science, Basic Energy Sciences (BES), under award DE-SC0019443 (D.R.G., J.-H.C., and X.X.). Some of the electron microscopy data were collected in the William R. Wiley Environmental Molecular Sciences Laboratory (EMSL), a national scientific user facility sponsored by DOE's Office of Biological and Environmental Research and located at Pacific Northwest National Laboratory. Some of the nanocrystal spectroscopy work was supported by the Northwest Institute for Materials Physics, Chemistry, and Technology (NW IMPACT). This research was additionally partially supported by an appointment to the Intelligence Community Postdoctoral Research Fellowship Program at UW (S.E.C.), administered by Oak Ridge Institute for Science and Education through an interagency agreement between the U.S. Department of Energy and the Office of the Director of National Intelligence. K.T.K. gratefully acknowledges support from the U.S. Department of Energy Office of Science Graduate Student Research (SCGSR) program for some of the TEM characterization. The SCGSR program is administered by the Oak Ridge Institute for Science and Education (ORISE) for the DOE. ORISE is managed by ORAU under contract number DE-SC0014664. Part of this work was conducted at the Molecular Analysis Facility, a National Nanotechnology Coordinated Infrastructure site at UW that is supported in part by the NSF (ECC-1542101), UW, the Molecular Engineering & Sciences Institute, the Clean Energy Institute, and the National Institutes of Health.

**Author Information**
[‡] These authors contributed equally to this work.
[∥] Present address: Department of Chemistry, Mississippi State University, Box 9573, Mississippi State, Mississippi 39762, USA
[§] Present address: School of Chemistry, Trinity College Dublin, The University of Dublin, College Green, Dublin 2, Ireland

**Contributions**
M.C.D., S.E.C., and D.R.G. conceived the experiments. S.E.C. and A.R. synthesized the nanocrystals and performed the analytical nanocrystal characterization. P.M., Q.J., and Z.L.

prepared the single-crystal sample, and P.M. and Q.J. performed the VSM magnetometry. M.C.D., K.T.K., and G.Z. performed and analyzed the electron microscopy measurements. M.C.D. and S.E.C. performed the optical and magneto-optical measurements and analyzed the data. All authors contributed to analysis of the results and preparation of the manuscript.

**Competing Interests**
The authors declare no competing financial interests.

**Supporting Information**
Additional synthesis discussion, TEM characterization of $CrBr_3$ nanoplatelets, derivative MCD magnetization *vs* magnetic field data, variable-temperature MCD spectra for $CrBr_3$ and $Cr(I_{1-x}Br_x)_3$ alloy nanoplatelets and discussion, estimation of $CrI_3$ nanoplatelet blocking temperatures.

**Table of Contents Graphic**

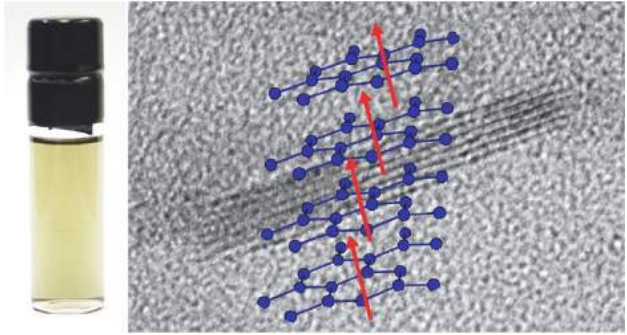

*Supporting Information for*

# 2D van der Waals Nanoplatelets with Robust Ferromagnetism

Michael C. De Siena,[1,‡] Sidney E. Creutz,[1,‡,||] Annie Regan,[1,§] Paul Malinowski,[2] Qianni Jiang,[2] Kyle T. Kluherz,[1,4] Guomin Zhu,[3,4] Zhong Lin,[2] James J. De Yoreo,[1,3,4] Xiaodong Xu,[2,3] Jiun-Haw Chu,[2] and Daniel R. Gamelin[1]*

[1]*Department of Chemistry, University of Washington, Seattle, WA 98195, USA*
[2]*Department of Physics, University of Washington, Seattle, WA 98195, USA*
[3]*Department of Materials Science and Engineering, University of Washington, Seattle, WA 98195, USA*
[4]*Physical Sciences Division, Pacific Northwest National Laboratory, Richland, WA 99352, USA*

*Corresponding author's e-mail: gamelin@uw.edu*

**S1: Additional description of the synthesis of CrI$_3$ nanoplatelets.** Attempts to synthesize CrX$_3$ using, *e.g.*, Cr(III) acetylacetonate, Cr(III) nitrate, or "Cr(III) acetate"[1] in carboxylate and amine surfactants either gave no readily isolable product, formed unidentified amorphous precipitates, or formed chromium oxide.[2] In most cases, we observed no reaction or even dissolution of these simple precursors until very high temperatures (>330 °C) were reached and sustained for extended periods.

By contrast, CrX$_3$ nanocrystals could be successfully prepared using Cr(OR)$_3$ (R = 1,1-di-*t*-butylethoxide) as the Cr(III) precursor. Cr(OR)$_3$ is a three-coordinate, highly air-sensitive complex of the bulky alkoxide "ditox" (2,2-di-*t*-butylethoxy) ligand, and it is highly soluble in alkane and arene solvents including pentane, toluene, and octadecene.[3] The anion source was trimethylsilyl halide (TMSX, X = I, Br). Cr(OR)$_3$ and TMSI do not undergo any observable reaction at room temperature or upon mild (<100 °C) heating, likely due to the considerable steric bulk of the reagents. Heating an anhydrous toluene solution of Cr(OR)$_3$ with excess Me$_3$SiX to 135 °C initiates nucleation of CrX$_3$ nanoplatelets. Reaction of the alkane-soluble iodide precursor TMSX with the alkoxide ligands of Cr(OR)$_3$ putatively releases the halide concomitant with formation of a silyl ether byproduct (Me$_3$SiOR). This silyl ether byproduct likely undergoes further decomposition *in situ*. The overall synthesis is summarized in Scheme S1.

**Scheme S1. Synthesis of CrX$_3$ (X = I, Br) nanoplatelets**

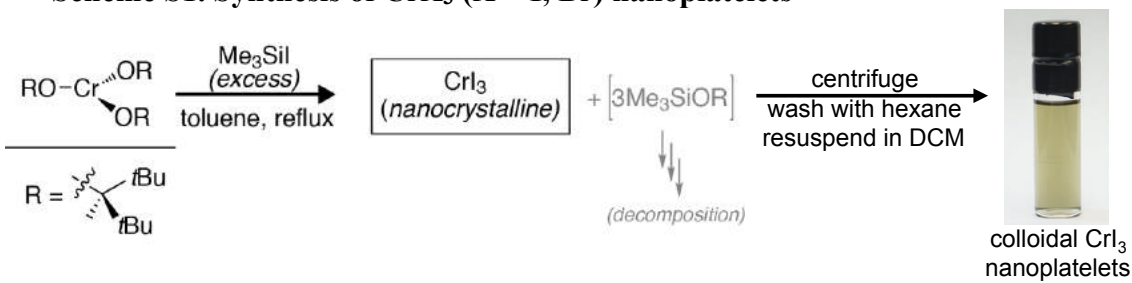

Figure S1 shows data from TEM measurements on CrBr$_3$ nanoplatelets. These data are similar to those presented for CrI$_3$ nanoplatelets in the main text.

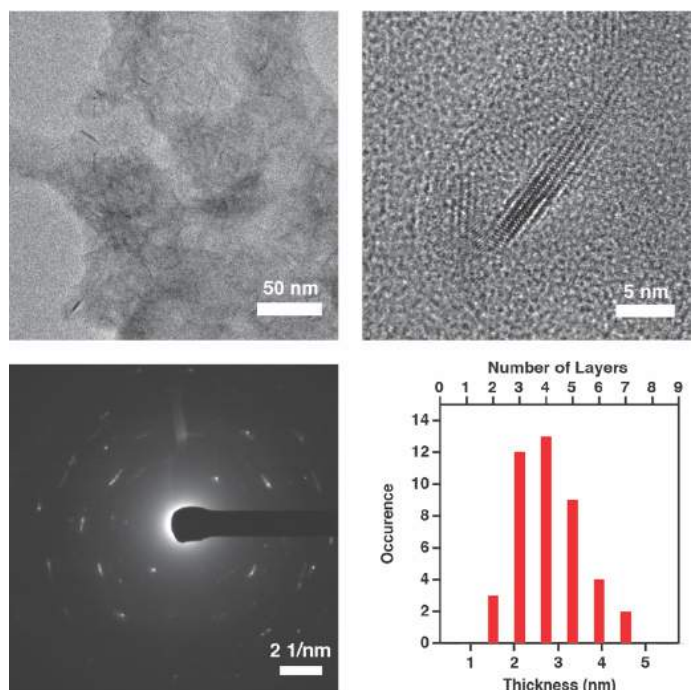

**Figure S1: Characterization of CrBr$_3$ nanoplatelets.** **(A)** TEM image of an aggregate of CrBr$_3$ nanoplatelets. **(B)** HR-TEM image of a CrBr$_3$ nanoplatelet lying on its short side. Lattice fringes are clearly seen. The nanocrystal in this image shows 5 individual CrBr$_3$ monolayers within the van der Waals nanoplatelet structure. **(C)** Selected area electron diffraction (SAED) image of aggregated CrBr$_3$ nanoplatelets. **(D)** Distribution of CrBr$_3$ nanoplatelet thicknesses, determined by measuring the sizes of >40 individual nanocrystals.

**S2: Inflections in magnetization data of CrI$_3$ nanoplatelets.** The data in the main text show multiple distinct inflections at intermediate fields during the MCD field sweep measurement of CrI$_3$ nanoplatelets. These inflections are more easily seen in the derivative of the field sweep data, as shown in Fig. S2.

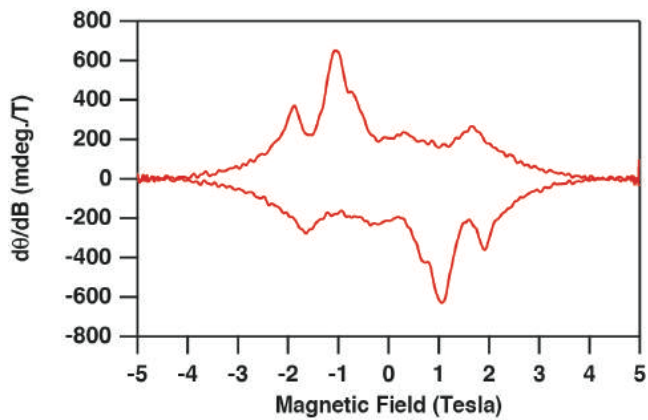

**Figure S2: Inflections in the MCD magnetization *vs* magnetic field data for CrI$_3$ nanoplatelets.** First derivative of the MCD magnetization field sweep data presented in the inset to Fig. 4b of the main text. The derivative highlights the inflections in the hysteresis loop. These inflections are assigned to spin flip events of individual CrI$_3$ layers.

**S3: Anion alloying and its effects on spectroscopy and $T_C$.** Cr(I$_{1-x}$Br$_x$)$_3$ nanoplatelets were prepared by mixing TMSI and TMSBr anion precursors during the synthesis. Varying the ratio of TMSI to TMSBr changes the ratio of Br$^-$ to I$^-$ incorporated into the nanoplatelets, and this ratio can be finely tuned across the entire range of $0 \leq x \leq 1$, but the nanoplatelet stoichiometry differs from the nominal (precursor) stoichiometry of the reaction. To illustrate, Fig. S3a plots the analytical I$^-$ contents measured in a series of Cr(I$_{1-x}$Br$_x$)$_3$ nanoplatelets as a function of the amount of TMSI added during synthesis. Halide compositions were determined by analysis of EDX data and by Vegard's law analysis of the (060) XRD reflection (inset of Fig. S3a); both approaches yield the same results. The elevated I$^-$ incorporation into the nanoplatelets is attributed to the greater reactivity of TMSI compared to TMSBr. These data demonstrate facile chemical control over the anion compositions in these colloidal CrX$_3$ nanoplatelets.

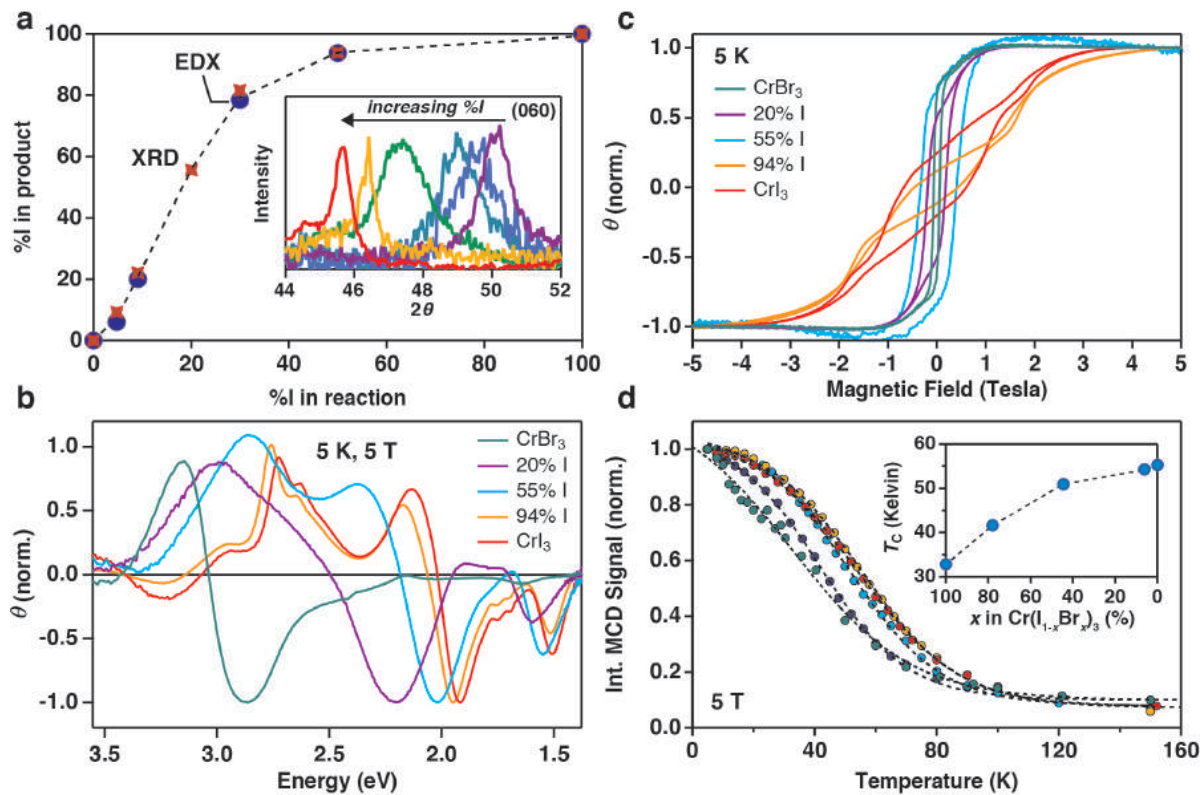

**Figure S3: Structural and magnetic characterization of alloyed Cr(I$_{1-x}$Br$_x$)$_3$ nanoplatelets. (A)** Measured mole percentage of iodide in the product for a given nominal iodide percentage in the reactants (TMSI and TMSBr). Compositions were determined by both XRD (Vegard's law, red crosses) and SEM/EDX (purple circles) measurements, which agree well. Inset: powder XRD data for the different alloy compositions, ranging from CrI$_3$ (red) to CrBr$_3$ (purple). The (060) reflection, typically the best resolved in the PXRD data, is shown. The dashed curve is a guide to the eye. **(B)** MCD spectra measured at 5 K and 5 T for different alloy compositions (given compositions are as determined by PXRD). **(C)** Field vs. magnetization sweeps for four different compositions of randomly oriented Cr(I$_{1-x}$Br$_x$)$_3$ nanoplatelets. **(D)** Integrated absolute MCD signal as a function of temperature for four different compositions of Cr(I$_{1-x}$Br$_x$)$_3$. Inset: Curie temperatures determined from the data in panel C and plotted as a function of iodide content for Cr(I$_{1-x}$Br$_x$)$_3$ nanoplatelets. The dashed curves are guides to the eye.

Anion alloying has dramatic effects on the nanoplatelet optical spectra and, in particular, on the stability of the ferromagnetic phase. Figure S3b plots 5 K, 5 T MCD spectra of a series of Cr(I$_{1-x}$Br$_x$)$_3$ nanoplatelets with $x$ ranging from 0 to 1. The energies of both the ligand-field and LMCT transitions redshift with decreasing $x$, as expected from the endpoint data in Fig. 3 of the main text. Figure S3c plots magnetic hysteresis data measured by MCD for the same samples. Increasing $x$ reduces the coercivity, narrowing the hysteresis. Whereas spin-flip inflections are clearly visible in the hysteresis data for the CrI$_3$ ($x$ = 0.00) and lightly bromide-doped ($x$ = 0.06) nanoplatelets, such features become less evident at higher $x$, where $B_{sat}$ is diminished. Nevertheless, even in the CrBr$_3$ nanoplatelets a small foot in the magnetization curve is seen at ~0.3 T that appears to stem from the same phenomenon. Figure S3d plots the integrated MCD intensity (absolute) as a function of temperature for the various Cr(I$_{1-x}$Br$_x$)$_3$ nanoplatelets. The spectra used for this analysis are presented in Fig. S4. $T_C$ values determined from these data are summarized in the inset of Fig. S3d, which shows that $T_C$ decreases as $x$ increases, reaching $T_C$ = 33 K at $x$ = 1 (CrBr$_3$). This value of $T_C$ is slightly smaller than the value (37 K) reported for bulk CrBr$_3$,[4] suggesting confinement effects in CrBr$_3$ similar to those found in the CrI$_3$ nanoplatelets.

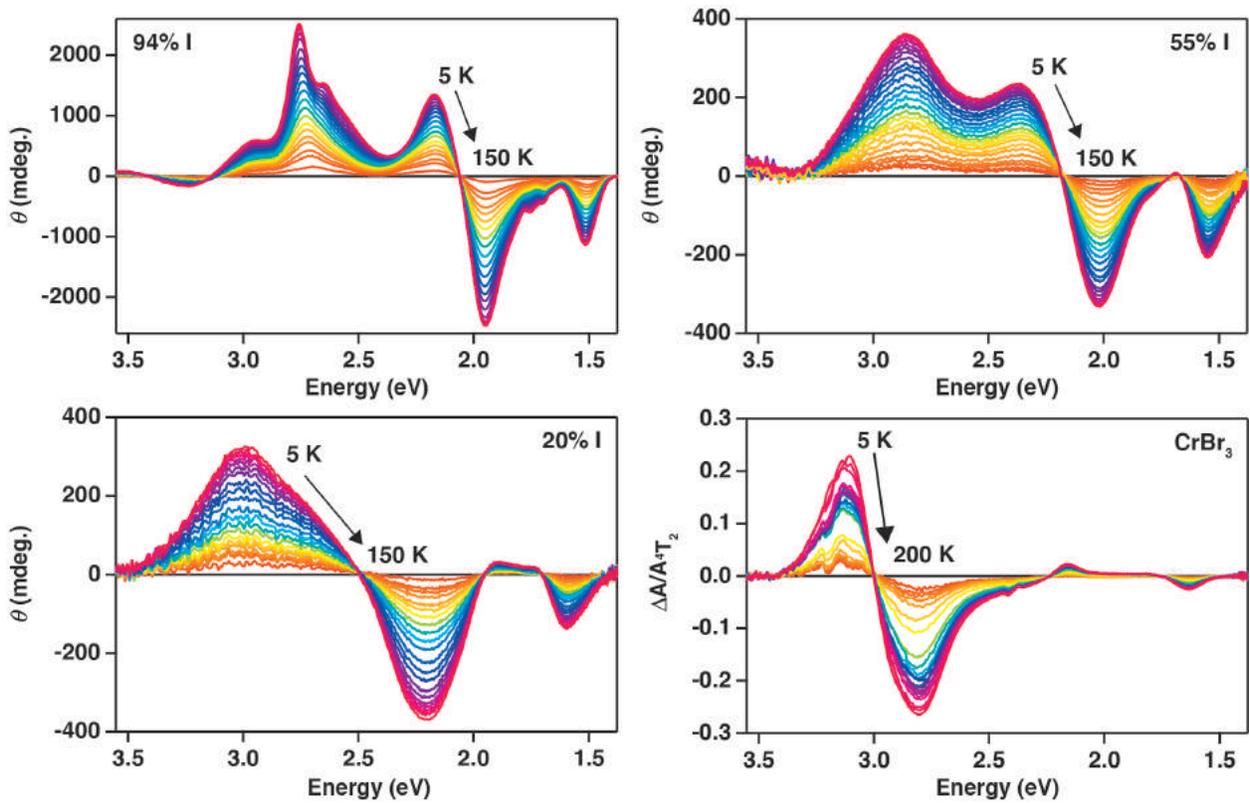

**Figure S4: Variable-temperature magnetic circular dichroism spectra of Cr(I$_{1-x}$Br$_x$)$_3$ nanoplatelets.** Temperature dependence (5 to 150 K) of the MCD spectra of Cr(I$_{1-x}$Br$_x$)$_3$ nanoplatelets, measured at 5 T. These spectra were used to determine the data points in Fig. S3d.

**S4: Estimation of the energy barrier to magnetization reversal in CrI$_3$ nanoplatelets.**
The Néel-Arrhenius equation (eq S1) describes the time constant for aligned spins to reverse their orientation:

$$\tau_N = \tau_0 e^{KV/k_B T} \quad (S1)$$

where $\tau_0$ is the attempt time, $K$ is the magnetocrystalline anisotropy constant, and $V$ is the particle volume. The product $KV$ represents the energy barrier to spin reversal of the single-domain particle. Idealizing the CrI$_3$ nanoplatelets as hexagons, we can express this energy barrier as:

$$E_{spin\ reversal} = K \cdot \frac{3\sqrt{3}}{8} d^2 h \quad (S2)$$

where $d$ is the platelet diagonal and $h$ is its height. $K$ has been measured for bulk CrI$_3$ where it is found to be temperature dependent.[5] Taking the values of $K$ = 50 kJ/m$^3$ (0.31 meV/nm$^3$) measured at ~$T_C$ and 300 kJ/m$^3$ (1.86 meV/nm$^3$) measured at low temperature, and using the interlayer spacing of 0.65 nm for the hexagon height, Fig. S5 plots $KV$ (meV) as a function of nanoplatelet size ($d$). For comparison, $k_B T_C$ is also plotted. This analysis shows that with these assumptions, $KV > k_B T_C$ for CrI$_3$ dimensions larger than $d \sim 6.5$ nm.

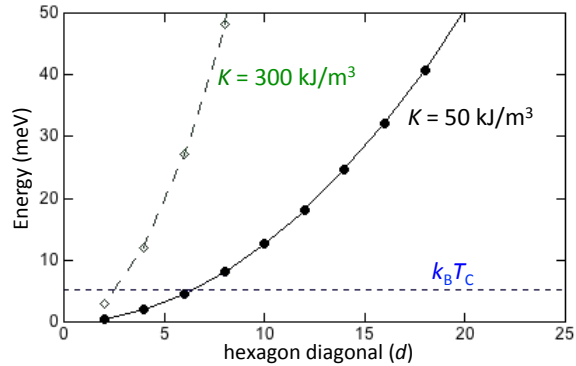

**Figure S5: Size dependence of the barrier to magnetization reversal in CrI$_3$ nanoplatelets.** Plots of estimated barriers to magnetization reversal ($KV$) for CrI$_3$ as a function of nanoplatelet size. Energy barriers were estimated from eq S2 for two different experimental (ref. 5) values of the volumetric anisotropy constant, $K$. $K$ = 50 kJ/m$^3$ (0.31 meV/nm$^3$) was measured at ~$T_C$, and $K$ = 300 kJ/m$^3$ (1.86 meV/nm$^3$) was measured in the low-temperature limit. For comparison, the value of $k_B T$ at $T_C$ is also included in the plot.

**Author Information**
[‡] These authors contributed equally to this work.
[∥] Present address: Department of Chemistry, Mississippi State University, Box 9573, Mississippi State, Mississippi 39762, USA
[§] Present address: School of Chemistry, Trinity College Dublin, The University of Dublin, College Green, Dublin 2, Ireland